\documentclass[aps, pra, preprint, groupedaddress, amsfonts,
               amsmath, amssymb, showpacs, nofootinbib]{revtex4-1}
\usepackage{microtype}
\usepackage{graphicx}
\usepackage{amssymb}
\usepackage{mathtools}
\usepackage{epstopdf}
\usepackage[utf8]{inputenc}
\usepackage[T1]{fontenc}
\usepackage[usenames,dvipsnames]{xcolor}
\usepackage{hyperref}

\usepackage{color}% \textcolor{#1}{#2}

\begin{document}
\title{Classical-trajectory model for ionizing proton-ammonia molecule collisions: the role of multiple ionization}

\author{Alba Jorge}  
\email[]{albamaria.jorge@gmail.com}
\affiliation{Departamento de Qu\'imica, Universidad Aut\'onoma de Madrid, Cantoblanco, E-28049 Madrid, Spain}

\author{Marko Horbatsch}  
\email[]{marko@yorku.ca}
\affiliation{Department of Physics and Astronomy, York University, Toronto, Ontario, Canada M3J 1P3}

\author{Tom Kirchner}  
\email[]{tomk@yorku.ca}
\affiliation{Department of Physics and Astronomy, York University, Toronto, Ontario, Canada M3J 1P3}
%%
%date{\today}
%
\begin{abstract}
We use an independent electron model with semi-classical approximation to electron dynamics to investigate 
differential cross sections for electron emission in fast collisions of protons with ammonia molecules.
An effective potential model for the electronic orbitals is introduced, and utilized in the context of the classical-trajectory Monte Carlo (CTMC)
approach for single-electron dynamics. Cross sections differential in electron emission angle and energy are compared with
experimental data. Compared to previous scattering-theory based quantum-mechanical results the
time-dependent semi-classical CTMC approach provides results of similar quality for intermediate and high ionized electron energies.
We find some discrepancies in the total cross sections for $q$-fold ionization 
between the present model and independent-atom-model calculations.  
The double ionization cross sections are considerably larger than recent experimental data which are derived from coincidence counting of 
charged fragments. The calculated triple ionization cross sections exceed the experimental coincidence data for $q=3$ by several orders
of magnitude at intermediate energies.
\end{abstract}

\maketitle
\section{Introduction}
\label{intro}
Collisions of protons and multi-charged ions with molecules has become an active field of research.
A number of approaches has been used for the water vapour target given its significance for radiotherapy.
Among the quantum-mechanical methods from stationary scattering theory are the Born approximation~\cite{Senger1988,Itoh13}, as well as 
versions of the continuum distorted wave (CDW) theory~\cite{Galassi_2012,Tachino_2015,Bhattacharjee_2016,Gulyas_2016,Purkait2017,Terekhin_2018,PhysRevA.108.032815}. Within the semi-classical approximation to the nuclear motion
there were attempts to solve the time-dependent Schr\"odinger equation (TDSE) within a mean-field approximation,
or density functional theory~\cite{hjl09,Murakami12a,Murakami12b}. 
A numerical solution of the TDSE with a three-center model potential was given in Ref.~\cite{ERREA2015}.
Another quantum-mechanical approach is the independent-atom model (IAM) where accurate
cross sections obtained in an independent-electron model (IEM) for collisions with constituent atoms are combined to form
cross sections for the given molecular target~\cite{Paredes15,hjl16,hjl20a,PhysRevA.106.022813}. The development of methods based on the TDSE to produce ionization
spectra, i.e., cross sections differential in electron energy and/or angle, has been slow, focusing so far mostly on collisions
with simple atomic targets or molecular hydrogen. Therefore, many studies resort to the classical treatment of electron motion, which can be viewed
as replacing the quantum Liouville equation by its $\hbar=0$ limit, i.e., reducing the electronic motion problem to classical statistical mechanics.

In the context of ion-molecule collisions the classical trajectory Monte Carlo (CTMC) methods have been developed for
molecular targets by a few groups~\cite{PhysRevA.83.052704,Sarkadi2015,PhysRevA.99.062701}. They can be realized at the level of an IEM, or pushed to an $N$-electron approach, although
the inclusion of the electron-electron interactions remains a big problem. Therefore, the $N$-electron calculations often remain
effectively close to single-electron models~\cite{Horbatsch_1992}, except that they allow to obtain multiple-electron event analysis directly without the
statistical approach required by the IEM analysis~\cite{Bachi_2019,PhysRevA.106.012808,atomsOtranto}.

The CTMC-IEM approach for ion-molecule collisions has recently been extended to the level of time-dependent mean-field theory in order to deal with highly
charged projectiles, and in order to be able to calculate capture processes correctly at lower energies~\cite{Jorge2020}.
Recently differential electron emission in $250 \ \rm keV$ proton-water molecule collisions were reported together with new experimental
data, as well as CDW theory (with eikonal initial state)~\cite{PhysRevA.105.062822}.

Our motivation for looking in detail at proton-ammonia collisions using the CTMC-IEM approach is as follows.
The molecular targets $\rm H_2O$, $\rm CH_4$, $\rm NH_3$ represent 10-electron systems with a central atom surrounded
by hydrogen atoms, and going from a planar towards different spatial geometries. Fast proton collisions with these targets
have been investigated experimentally, where the focus is on the charged fragments produced in these collisions. Recently emphasis
was placed on multiple ionization events, and some serious discrepancies were found in the case of the ammonia target~\cite{Wolff20}.
In contrast to the case of water vapor~\cite{Luna07,Tavares_2015}, where multiple ionization is found to be a significant contributor, 
and the case of methane~\cite{Luna19} where double ionization was clearly identified, the experimental ammonia results
did not support the case of a direct two-electron ionization process.

A recent analysis of the total multiple ionization cross sections for the three collision systems within an IAM approach~\cite{PhysRevA.106.022813}
highlights the controversy from one theoretical point of view. Here we would like to address the role of multiple ionization both
in electron emission (where the distinction between net (total) and single ionization is sometimes blurred in the literature), and
also from a total cross section point of view. We construct a potential model for the $\rm NH_3$ molecule 
in analogy to what was done for the water molecule~\cite{PhysRevA.83.052704,ERREA2015,PhysRevA.99.062701} and then apply
the CTMC approach within the IEM.

The layout of the paper is as follows. We begin in Sec.~\ref{sec:theory} 
with a short summary of the CTMC-IEM. 
For the $\rm p-NH_3$ system at $250 \ \rm keV$ collision energy and higher
we found the time-dependent mean-field model~\cite{Jorge2020} to give practically the same
results for differential electron emission as the static CTMC model, and thus we focus on the latter in the present work. 
We provide in Sec.~\ref{sec:ctmc-iem} also the expressions for doubly differential cross sections (DDCS)
in ejected electron energy and direction for both the typical case of net ionization, and 
for the specific case of single ionization.
In Sec~\ref{sec:potmod} we discuss the construction of the potential model.
Results are presented and compared with experimental data and 
selected previous calculations
in Sec.~\ref{sec:results}, first for
differential electron emission (in Sec.~\ref{sec:ddcs}) and then
for total cross sections (in Sec.~\ref{sec:tcs}). The paper ends with
conclusions in Sec.~\ref{sec:conclusions}.
Atomic units, characterized by $\hbar=m_e=e=4\pi\epsilon_0=1$, are used unless otherwise stated.

% more on molecules, need for data?
% look at Lokesh's piece for ICPEAC roadmap (nothing much there; some emph on diatomics)
% two layers of statistics: classical - quantum and one electron - many electrons

\section{Theory}
\label{sec:theory}
We provide a brief summary of the CTMC-IEM with static target potential, as developed for the water
molecule in Ref.~\cite{PhysRevA.99.062701}. Within the semiclassical impact parameter approximation the proton trajectories
follow straight lines, and provide an explicitly time-dependent potential for the electronic orbitals bound in a
multi-center potential which is described in the next subsection.
The total Hamiltonian of the system is a sum of single-particle Hamiltonians for the electronic orbitals, i.e., $H=\sum_j{h_j(t)}$,
with orbital-independent $h_j(t) \equiv h(t)$.
For reference we use the single-center Hartree-Fock orbital energies of Moccia~\cite{Moccia_64b}:
$\epsilon_{1a_1}\approx -15.52$, $\epsilon_{2a_1}\approx -1.12$, $\epsilon_{3a_1}\approx -0.415$, and $\epsilon_{1ex}=\epsilon_{1ey} \approx -0.596$ (given in Hartree
units), and spin degeneracy leads to double occupation.

\subsection{CTMC-IEM approach}
\label{sec:ctmc-iem}

The single-particle Hamiltonians for the evolution of orbitals can be written as
\begin{equation}
	h(t) =  \frac{{\bf p}^2}{2} + v_{\rm mod}({\bf r}) -\frac{Z_p}{|{\bf r}-{\bf R}(t)|}.
\label{eq:hamilton}
\end{equation}
Here $Z_p=1$ for protons and ${\bf R}(t)$ is the (straight-line) proton trajectory. The model potential $v_{\rm mod}({\bf r})$ is evaluated for a random
orientation of the molecule with ensemble averaging implied. For the fast collisions considered in this work neither vibrational nor rotational motion
of the target molecule is taken into account. At this point one has two options: {\it (i)} quantum mechanical evolution of the time-dependent orbitals (cf. Ref.~\cite{ERREA2015});
 {\it (ii)} CTMC calculations in the spirit of the $\hbar=0$ approximation. We follow the latter approach in this work.

The electronic orbitals are simulated by ensembles of classical trajectories which satisfy Hamilton's equations. In contrast to the time-dependent
mean-field approach of Ref.~\cite{Jorge2020}, which was also applied for the present work, but which did not yield significantly
different results, the methodology is then straightforward CTMC in the impact parameter approximation.
Ensembles of electron trajectories representing MOs are analyzed both for net and for multiple differential electron emission.

Initial conditions for the trajectories are based on the microcanonical distribution. In  quantum simulations one
has to make sure that the orbitals are placed in eigenstates of the target Hamiltonian. In the classical Liouville approach, however,
one is not forced to apply such a condition~\cite{RevModPhys.55.245}.
The $\hbar=0$ approximation to the quantum Liouville equation requires for stable evolution only that the initial distribution be a pure function of 
energy~\cite{Cohen_1985}.
Within the microcanonical approach
one constructs stable initial orbital distributions for a given external potential by combining a chosen initial value
for the orbital energy with a distribution over angular momenta consistent with the microcanonical ensemble~\cite{Abrines_1966}.
Examples of such classical microcanonical distributions were described, e.g,
for the water molecule in Ref.~\cite{PhysRevA.83.052704} and for the uracil molecule in Ref.~\cite{Sarkadi2015}.

The analysis whether a trajectory contributes to ionization or capture after proton and molecule are separated by a large distance is as follows:
at the final proton-molecule distance of about 500 atomic units the energy of the test electron representing an MO is calculated with respect
to projectile and target. If both energies are positive the event is counted as ionization, and the information about ionized electron energy
and direction is recorded.

The number of trajectories is rather large due to the sampling of different impact parameters and random molecular orientations.
The sampling is required since the experimental data are insensitive to this information, i.e., projectile deflections or target recoil
motion are not recorded. Most works in the literature report net differential ionization cross sections by summing over the contributions
from all MOs. The single probability for ionization from orbital $j$ is defined in terms of the number of ionizing trajectories $n_j^{\, \rm ion}$ as 
\begin{equation}
	p_j^{\, \rm ion} = \frac{n_j^{\, \rm ion}}{n_{j,{\rm tot}}} .
\label{eq:psingle}	
\end{equation}
A binning procedure allows to obtain such a single probability differential in electron energy and scattering angles.
The net differential cross section
can be written as
\begin{equation}
	\frac{{\rm{d}}^2P_{\rm net}^{\rm ion}}{{\rm{d}} E_{\rm{el}}{\rm{d}}\Omega_{\rm{el}}} =   
	2 \sum_{j=1}^m \frac{{\rm{d}}^2p_{j}^{\rm ion}}{{\rm{d}} E_{\rm{el}}{\rm{d}}\Omega_{\rm{el}}} .
\label{eq:pdiffnet2}
\end{equation} 
The interpretation of this expression is that an electron has been recorded with a given energy and direction (obviously with finite resolution)
independent of how many electrons were ionized overall. The factor of two accounts for spin degeneracy in each MO.
This analysis corresponds to the typical experimental situation.

Expressions for differential cross sections in the IEM specific for $q$-fold ionization were derived in Ref.~\cite{PhysRevA.99.062701}.
For $q=1$ we quote
\begin{equation}
	\frac{{\rm{d}}^2P_{q=1}^{\rm ion}}{{\rm{d}} E_{\rm{el}}{\rm{d}}\Omega_{\rm{el}}} =  2 \sum_{j =1}^m{\frac{{\rm{d}}^2 p_{j}^{\rm ion}}{{\rm{d}} E_{\rm{el}}{\rm{d}}\Omega_{\rm{el}}} (1-p_{j}^{\rm ion})
\prod_{k \ne j}^m {(1-p_k^{\rm ion})^2}} .
\label{eq:pdq1}
\end{equation}
This expression describes events where an electron has been recorded at given energy and direction, while no other electrons were found in the continuum.
We report in Section~\ref{sec:ddcs} results for both cases, i.e., Eqs.~(\ref{eq:pdiffnet2}) and~(\ref{eq:pdq1}).

\subsection{Model potential}
\label{sec:potmod}
The target potential is modelled as a
sum of central potentials for each atom of the molecule:
\begin{equation}
v_{\rm mod} = v_{\rm N}(r_{\rm N}) + \sum_{i=1}^3 v_{\rm H}(r_{{\rm H}_i}) .
\label{eq:nh3pot1}
\end{equation}
Here $r_{\rm N}$ and $r_{{\rm H}_i}$ are the distances from the
active electron to the nitrogen and the $i=1,2,3$ hydrogen nuclei of the
NH$_3$ molecule. We follow the nuclear geometry obtained in single-center
self-consistent field (SCF) calculations~\cite{Moccia_64b}: the
N-H bond lengths are 1.928 a.u., the polar angles of the H atoms 
are 108.9$^\circ$ and the azimuthal angles are given as 90$^\circ$, 210$^\circ$, and 330$^\circ$, respectively.
The central potentials in Eq.~(\ref{eq:nh3pot1}) are assumed to take the forms
\begin{equation}
\begin{alignedat}{1}
&v_{\rm{N}}(r_{\rm{N}})=-\frac{7-N_{\rm{N}}}{r_{\rm{N}}}-\frac{N_{\rm{N}}}{r_{\rm{N}}}(1+\alpha_{\rm{N}}r_{\rm{N}})\exp(-2\alpha_{\rm{N}}r_{\rm{N}}) , \\
&v_{\rm{H}}(r_{\rm{H}})=-\frac{1-N_{\rm{H}}}{r_{\rm{H}}}-\frac{N_{\rm{H}}}{r_{\rm{H}}}(1+\alpha_{\rm{H}}r_{\rm{H}})\exp(-2\alpha_{\rm{H}}r_{\rm{H}}) , \\  
\end{alignedat}
\label{eq:nh3pot2}
\end{equation}
where $N_{\rm{N}} = 6.2775$, $\alpha_{\rm{N}}=1.525$, $N_{\rm{H}} = 0.9075$, and $\alpha_{\rm{H}}=0.6170$. These choices were made in the following way: 
The parameters $N_{\rm{H}}, \alpha_{\rm{H}}$ for the hydrogen atoms are the same as in previous
work for H$_2$O in which an analogous model potential was used~\cite{ERREA2015,PhysRevA.83.052704,PhysRevA.99.062701,PhysRevA.105.032814}.
The asymptotic large-$r$ behavior of the model potential should be $-1/r$. 
This fixes the screening charge parameter of the nitrogen atom to
$N_{\rm{N}}=9-3 N_{\rm{H}}=6.2775$. The remaining parameter $\alpha_{\rm{N}}$ was determined such that
the energy eigenvalues of the Hamiltonian $\frac{{\bf p}^2}{2} + V_{\rm mod} $ are in reasonable 
agreement with the SCF results from~\cite{Moccia_64b}. 
% With $\alpha_{\rm{N}}=1.525$ we obtain ... for the occupied MOs.
% Alba uses Moccia's eigenvalues, how about geometry?

%%%  maybe I have to spell out the potential and the parameters?  Not good

\section{Results}
\label{sec:results}

\subsection{Doubly-differential cross sections}
\label{sec:ddcs}

\begin{figure}
\begin{center}
\resizebox{0.8\textwidth}{!}{\includegraphics{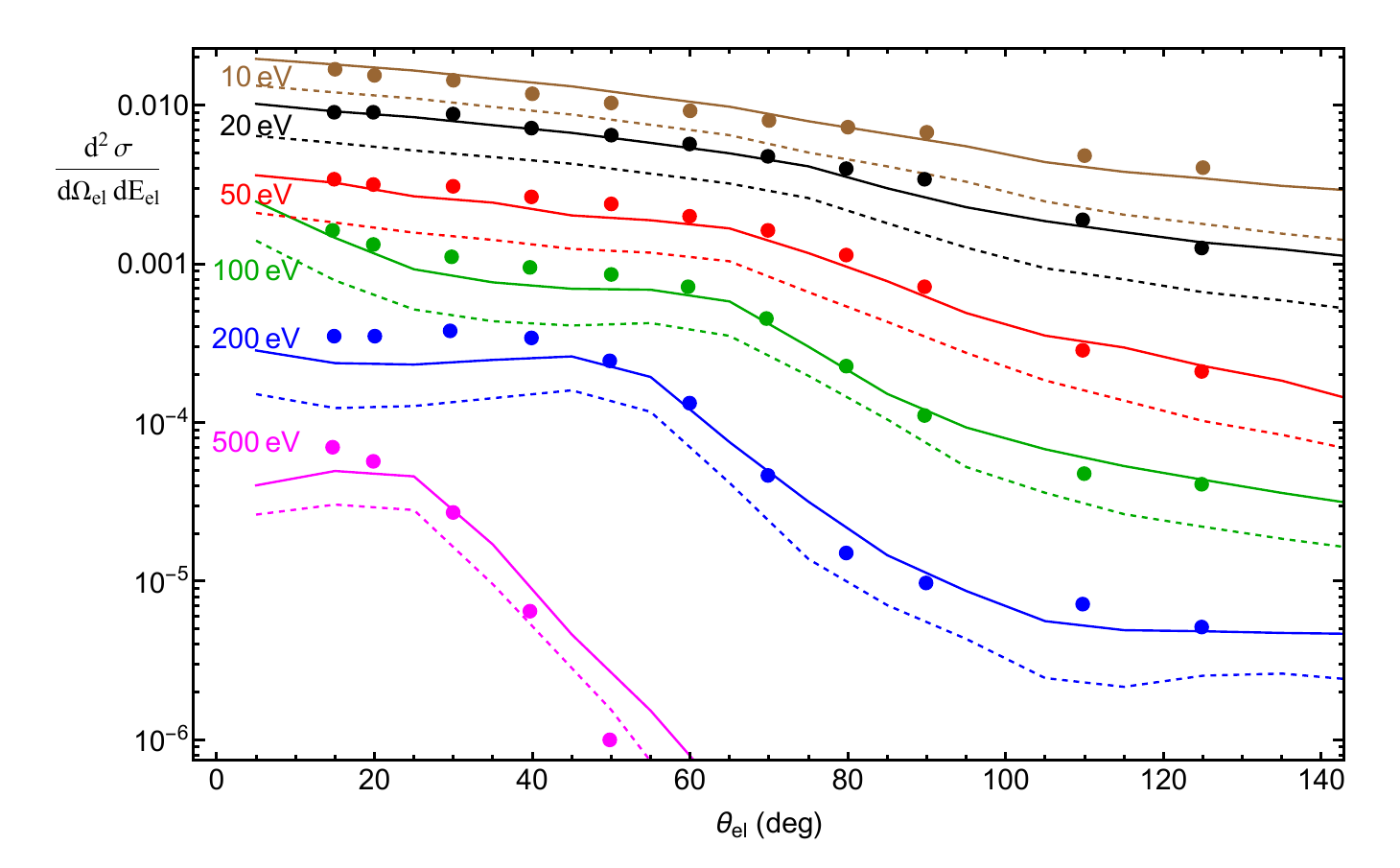}}
\vskip -0.5 truecm
\caption{%
Doubly differential cross section in units of {\AA}$^2/({\rm eV \, srad})$ for proton collisions with ammonia molecules at a collision energy of $E_{\rm p}=250 \ \rm keV$
for ionized electron energies of $E_{\rm el}=10, \,20,\, 50,\, 100,\, 200,\, 500 \ \rm eV$. Solid lines: present CTMC net ionization results obtained with Eq.~(\ref{eq:pdiffnet2}); 
dashed lines are for single ionization obtained with Eq.~(\ref{eq:pdq1}). 
The data points are the experimental results of Ref.~\cite{Lynch1976}, as reported in Refs.~\cite{Senger1988,Purkait2017}.
}
\label{fig:ddcs250}   
\end{center}
\end{figure}

In Fig.~\ref{fig:ddcs250} we show DDCS results for an impact energy of $E_{\rm p}=  250 \ \rm keV$.
The $\hbar=0$ approach to electron dynamics should work well at not-too-low ionized electron energies,
and we find that this is the case for all experimentally observed cases.
The net ionization results (solid lines, calculated with Eq.~(\ref{eq:pdiffnet2})) agree well with the experimental 
data with some factor-of-two deviations at $E_{\rm el}=200$ eV electron energies and forward angles.

One of the major objectives of this work is to illustrate the role of multi-electron processes within the CTMC-IEM.
The dashed lines show the results for single ionization, i.e., an electron is detected at given energy and angle, and it is the
only electron in the continuum. We observe that the results are of very similar shape and typically reduced by a factor of about two.
This implies that at 250 keV impact energy double (or higher multiple) ionization makes roughly  an equal contribution to the net
ionization cross sections. This is perhaps overlooked at times, based on the thought that protons should predominantly lead to
single ionization only. The ammonia molecule is an extended object and has a number of electrons that are bound on the scale
of an atomic ground-state hydrogen electron.

The experimental data were explained previously using quantum-mechanical stationary scattering theory methods,
which work well for low-energy electron emission.
Senger~\cite{Senger1988} used a plane-wave Born approach adopted from proton-atom scattering to
represent MOs as linearly combined atomic orbitals. Results were reasonable for ionized electron energies above
20 eV with some noticeable shortfall for backward scattering, particularly for $E_{\rm el}=200 \ \rm eV$.
Mondal~{\it et al.}~\cite{Purkait2017} extended a three-Coulomb-wave model and used the
single-center SCF orbitals of Moccia~\cite{Moccia_64b} to describe the initial MOs.

Tachino~{\it et al.}~\cite{Tachino_2015} extended the theory to the post and prior versions of the 
CDW-EIS model (EIS=eikonal initial state). They used the Moccia orbitals, as well as a 
linearly-combined atomic orbital SCF representation, both giving very close results at the intermediate
impact energy of $E_{\rm p}=250 \, \rm keV$. The post and prior
versions of the CDW-EIS model did show, however, remarkable differences.
There is some discussion in the literature as to which form ought to be favoured, and the situation may actually
depend on the collision energy. 
The prior version appears to have been favoured (e.g., commented upon in Ref.~\cite{PhysRevA.99.062701}).

\begin{figure}
\begin{center}
\resizebox{0.95\textwidth}{!}{\includegraphics{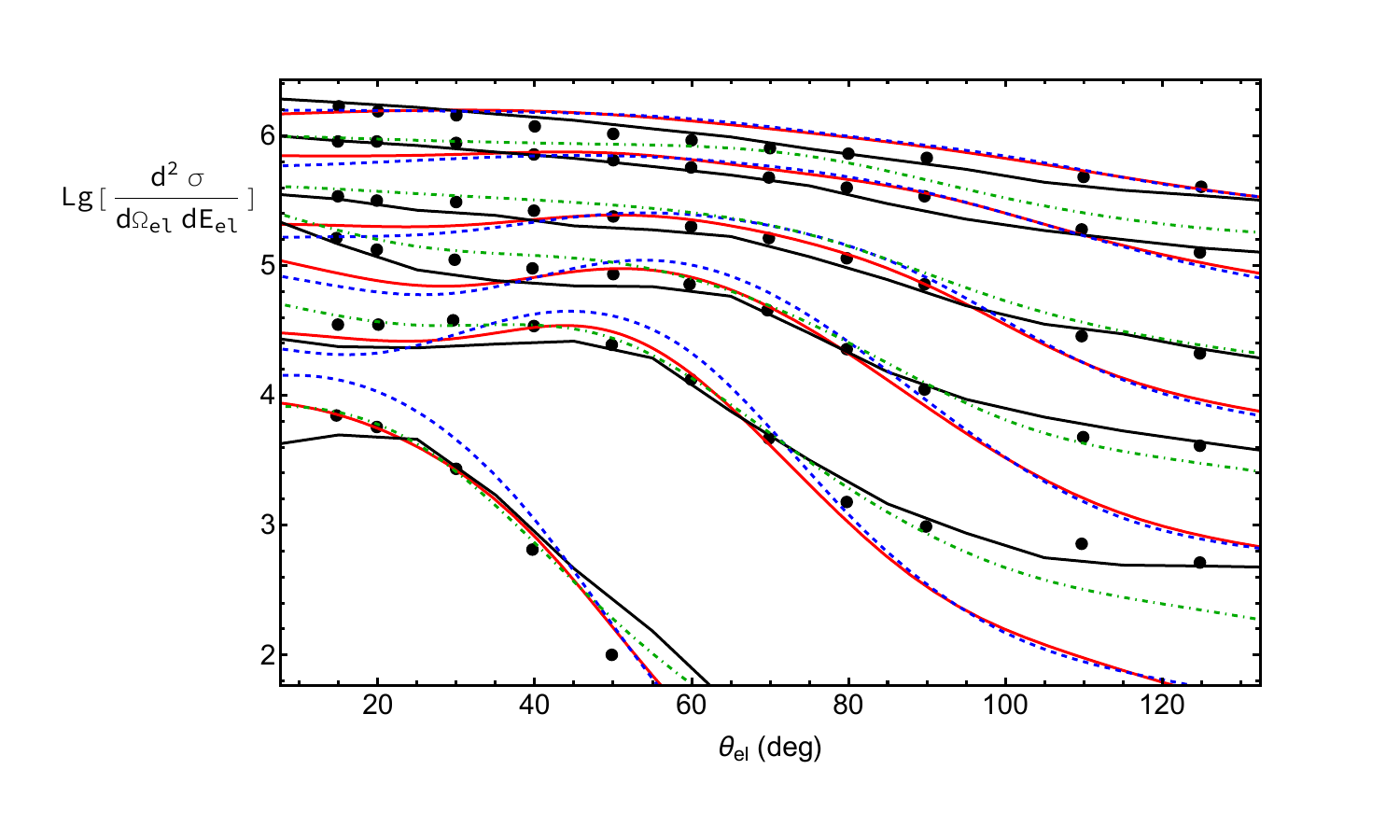}}
\vskip -0.5 truecm
\caption{%
Logarithm of the net doubly differential cross section for proton collisions with ammonia molecules at a collision energy of $E_{\rm p}=250 \ \rm keV$ as
compared to the CDW-EIS results of Tachino~{\it et al.}~\cite{Tachino_2015}
for ionized electron energies of $E_{\rm el}=10, \,20,\, 50,\, 100,\, 200,\, 500 \ \rm eV$ (from top to bottom). 
Solid black lines connecting binned data: present CTMC net ionization results obtained with Eq.~(\ref{eq:pdiffnet2}); solid red curves:
CDW-EIS (prior), dashed blue curves: CDW-EIS (post); green dash-dotted curves (starting at $E_{\rm el}=20 \, \rm eV$): three-Coulomb wave model of Ref.~\cite{Purkait2017}.
The value of 2 on the $y$-axis corresponds to a cross section value of $10^{-6}$
{\AA}$^2/({\rm eV \, srad})$.
}
\label{fig:tachino}   
\end{center}
\end{figure}

In Fig.~\ref{fig:tachino} we compare the results for net ionization from Fig.~\ref{fig:ddcs250} with these CDW-EIS results
and with the three-Coulomb-wave model cross sections of Ref.~\cite{Purkait2017}.
The post and prior forms provide very close results at low electron energies, and display some undulatory behaviour
at $E_{\rm el}>100$ eV and forward angles where the two models make somewhat different predictions. 

For $E_{\rm el}=20 \, \rm eV$ the CTMC results agree
with the experimental data which are also described reasonably well by three-Coulomb-wave model
of Ref.~\cite{Purkait2017} (which overestimates them a bit) and the CDW-EIS results of Ref.~\cite{Tachino_2015}
which fall a bit short at the largest angles.

At higher electron energies the correspondence principle allows the CTMC model to demonstrate its strength, since
it employs a more realistic multi-center interaction. For emission angles $\theta_{\rm el} \ge 30$ degrees it follows the data
very well, while the CDW-EIS results show some weakness in the backward direction.  The present results for $E_{\rm el}=50 \, \rm eV$
agree in shape with the results of Ref.~\cite{Purkait2017}.

For the CDW-EIS models the shortfall in predicted backward electron emission 
leads to an underestimation of the doubly differential cross section by about an order of magnitude
at the higher electron energies of 100 and 200 eV, while the CTMC results show very good agreement with the experimental data.
The good agreement of the CTMC-IEM results at backward angles and high electron energies is gratifying
and lends credibility to the experimental data. It is likely caused by the multi-center potential used in the CTMC approach
which results in a noticeable effect even after orientation-averaging of the molecule.

The three-Coulomb-wave model gives good results for 100 eV, and falls short by about a factor
of two at 200 eV when compared to experiment at the largest measured scattering angles of 110 and 125 degrees.
While we do not show the plane-wave Born data of Ref.~\cite{Senger1988} we note that the comparison
is provided in Fig.~7 of Ref.~\cite{Purkait2017}. Overall, the  three-Coulomb-wave model agrees with
the model used in Ref.~\cite{Senger1988}: a small relative enhancement of backward electron emission
at high energies appears to be a notable difference.

We note that in the case of $\rm p-H_2O$ collisions at $E_{\rm p}=250 \ \rm keV$ improvements to the CDW-EIS model
have been introduced which address the problem at backward angles~\cite{PhysRevA.105.062822}.
Concerning the comparison of the two versions of CDW-EIS we observe that the post form shows stronger undulations in the
cross sections for $E_{\rm el} \ge 100 \, eV$ and forward to intermediate scattering angles which are not supported by the data.
The prior form, on the other hand, fares better in this respect.

%\begin{figure}
%\begin{center}$
%\begin{array}{cc}
%\resizebox{0.48\textwidth}{!}{\includegraphics{DDCS250keV.pdf}}&
%\resizebox{0.48\textwidth}{!}{\includegraphics{DDCS1MeV.pdf}}

%\end{array}$
%\caption{
%}
%\label{fig:fig1}
%\end{center}
%\end{figure}

\begin{figure}
\begin{center}
\resizebox{0.7\textwidth}{!}{\includegraphics{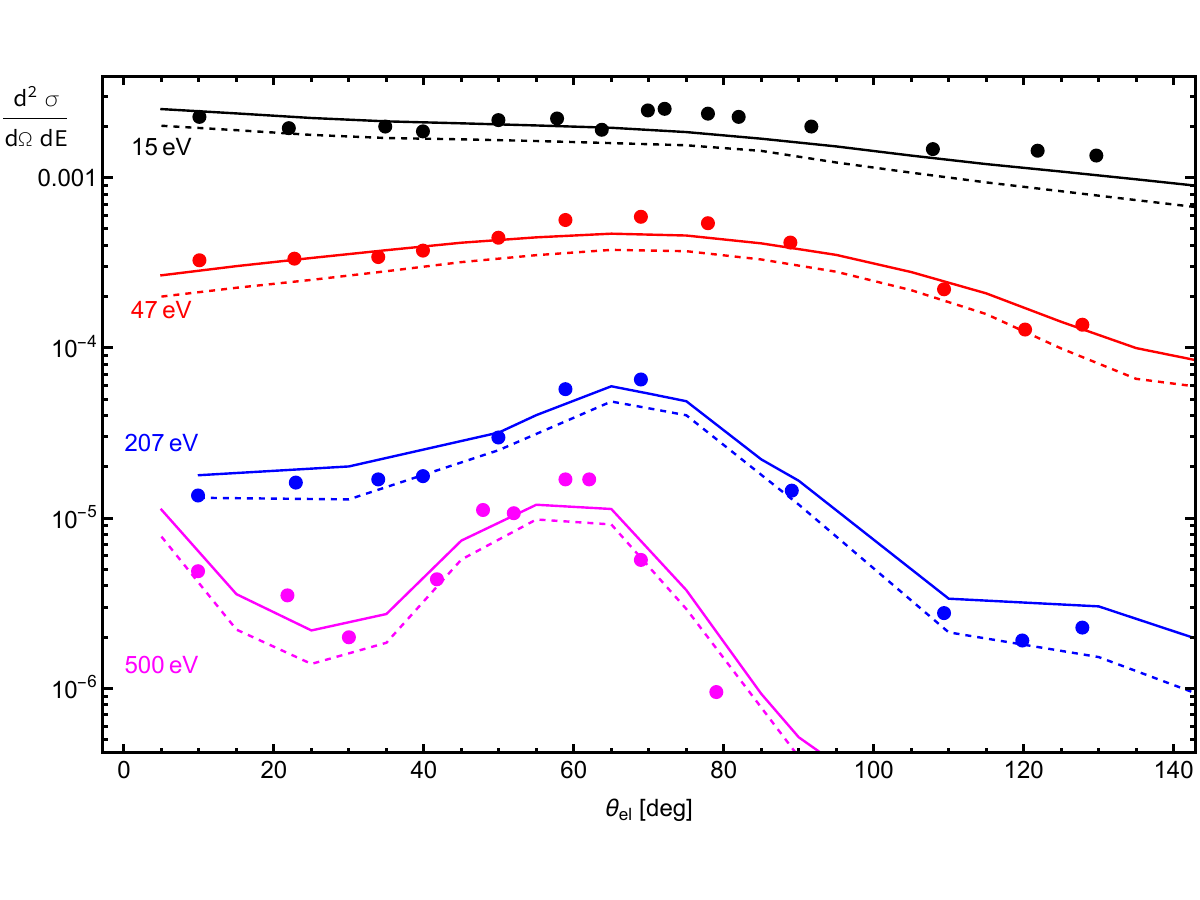}}
\vskip -0.5 truecm
\caption{%
Same as in Fig.~\ref{fig:ddcs250}, but for a collision energy of $E_{\rm p} =1 \ \rm MeV$.
The ionized electron energies are $E_{\rm el}=15,\, 47,\, 207,\, 500 \ \rm eV$.
The data points are the experimental results of Ref.~\cite{Lynch1976}, as reported in Ref.~\cite{Senger1988,Purkait2017}.
}
\label{fig:ddcs1000}   
\end{center}
\end{figure}

The case of 1 MeV impact energy is presented in Fig.~\ref{fig:ddcs1000}. A marked difference is found in the angular dependence of the DDCS.
The results are of similar quality as those of Mondal~{\it et al.}~\cite{Purkait2017} (not shown).
In some way the result for net ionization just confirms that the CTMC-IEM works at this high energy. 
We observe that multiple ionization plays a lesser role compared to 250 keV proton impact.
The single-ionization DDCS shown by the dashed lines follow the net ionization more closely in Fig.~\ref{fig:ddcs1000} than in Fig.~\ref{fig:ddcs250}.

The ratio of single to net DDCS varies with collision energy.
At 250 keV impact energy this ratio varies between 0.65 at forward angles and 0.5 at backward angles for low ionized electron energies.
For electron energies of 100 eV and higher it is closer to 0.5 and more uniform. For the impact energy of 1 MeV the ratio of single to net 
DDCS values moves towards $0.8-0.9$. It is at the higher end for low ionized electron energies and closer to the lower bracket at energies of
50 eV and higher with small variation with respect to emission angles.

Concerning the absolute height of the net ionization DDCS we note the good agreement between the CTMC-IEM and experimental results.
The 1 MeV collision energy is within the range where the deviation between the classical and quantum total ionization cross section behaviour 
is not yet noticeable (the difference being a logarithmic dependence missing in the classical case).

\begin{figure}
\begin{center}
\resizebox{0.7\textwidth}{!}{\includegraphics{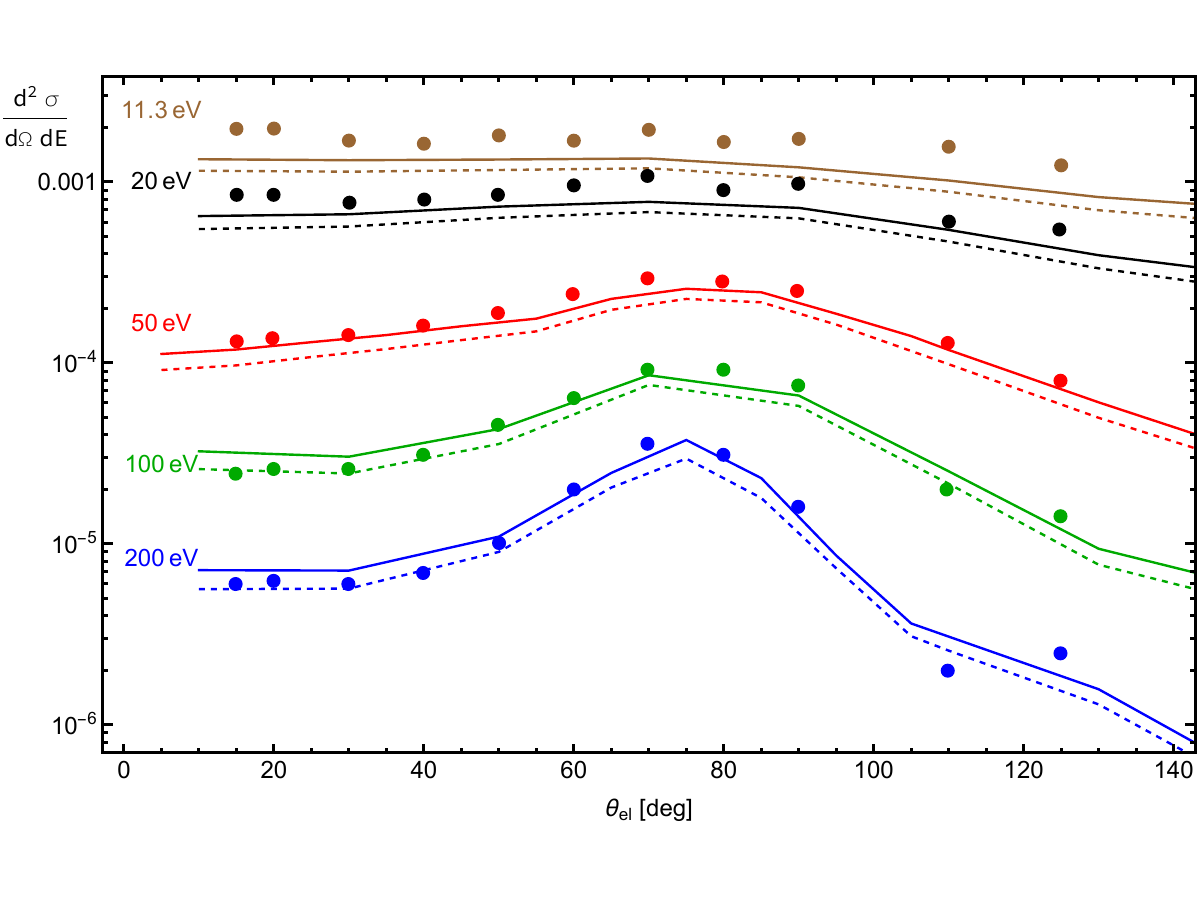}}
\vskip -0.5 truecm
\caption{%
Same as in Fig.~\ref{fig:ddcs250}, but for a collision energy of $E_{\rm p} = 2 \ \rm MeV$.
The ionized electron energies are $E_{\rm el}=11.3, \,20,\, 50,\, 100,\, 200 \ \rm eV$.
The data points are the experimental results of Ref.~\cite{Lynch1976}, as reported in Refs.~\cite{Purkait2017}.
}
\label{fig:ddcs2000}   
\end{center}
\end{figure}

In Fig.~\ref{fig:ddcs2000} we show results for the impact energy of 2 MeV.
We have not included data for the ionized electron energy of 1,000  eV, due to statistical 
limitations in the present calculations. For intermediate to high electron energies (50-200 eV) the comparison with the data is very good.
The contribution of single ionization towards the net DDCS is at the $80 \, \%$ level with small dependence on the emission angle and energy. 

For the lower electron energies of 20 eV, and particularly 11.3 eV the CTMC-IEM results clearly fall short of the experimental data.
This is likely to be caused by the lack of a dipole ionization mechanism in the semi-classical $\hbar=0$ approach, which is discussed in the literature,
cf. Ref.~\cite{Reinhold_1993}. One can then raise the question of the total cross section at such a high collision energy: does this shortfall
of low-energy electron production lead to an underestimation of the net ionization cross section? This is addressed in Section~\ref{sec:tcs}.
We note, however, that it is difficult to estimate the total cross section from the experimental data shown in Fig.~\ref{fig:ddcs2000},
since an extrapolation to low electron energies would be required. In turn, this also raises the question of absolute normalization of the
experimental DDCS data.

As mentioned above, the role of double (or higher multiple) ionization is reduced at higher collision energies, i.e., there is no longer a factor-of-two discrepancy
between the solid and the dashed lines as one compares the collision energies of 250 keV with 1 and 2 MeV. 
Nevertheless, the main conclusion from the analysis of CTMC data is that multiple ionization contributes significantly to the net DDCS data: it is quite important
at the lower energies. Another view of the situation can be obtained by comparing integrated (total) cross sections as a function of energy to see how
the net cross section is made up of single and multiple ionization contributions.

\subsection{Total cross sections}
\label{sec:tcs}

\begin{figure}
\begin{center}
\resizebox{0.8\textwidth}{!}{\includegraphics{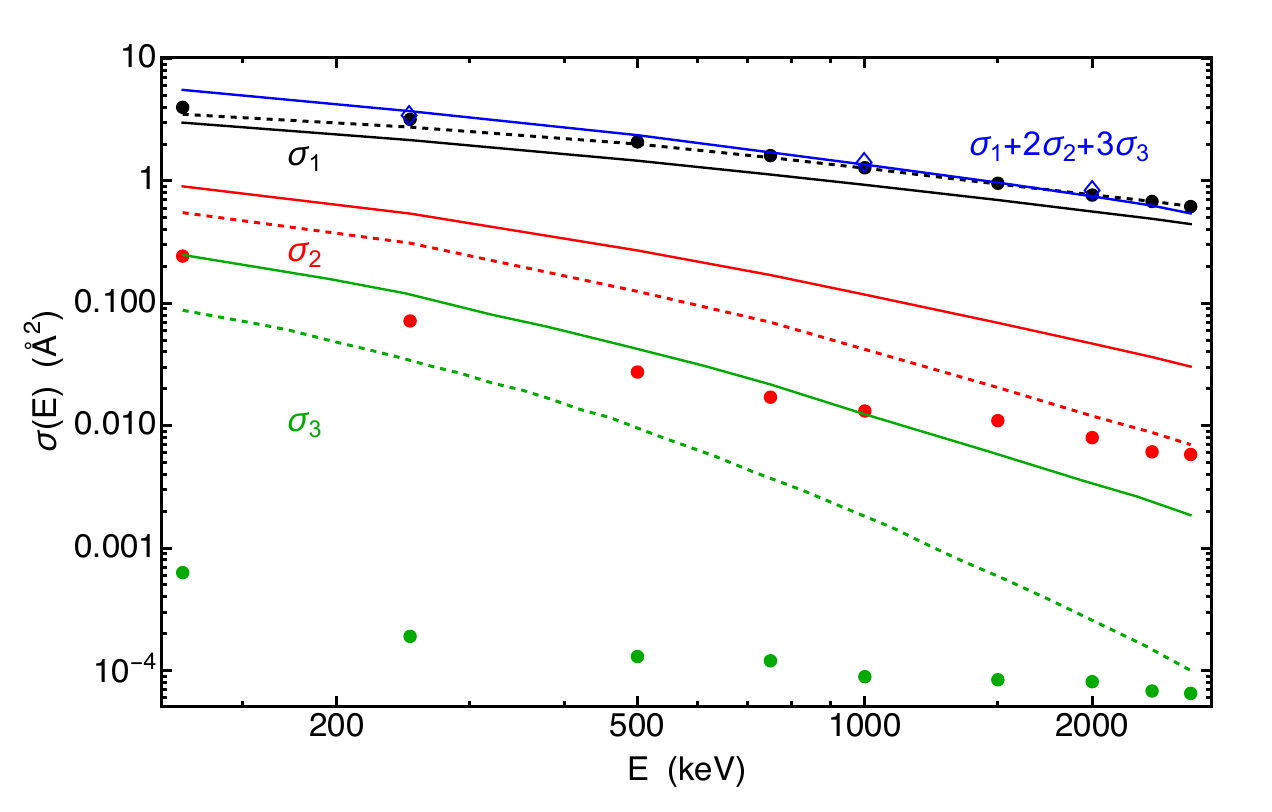}}
\vskip -0.5 truecm
\caption{%
Total cross sections (in units of {\AA}$^2$) for proton collisions with ammonia molecules as a function of  collision energy. 
Solid blue line: present CTMC approximate net ionization result, %$\sigma_1+2 \sigma_2+3 \sigma_3$
$\sum_{q=1}^3{q \, \sigma_q}$;
blue open diamonds: net ionization cross sections from Lynch~{\it et al.}~\cite{Lynch1976}.
Solid lines (black, red, green): present CTMC results for $\sigma_q$ with $q=1,2,3$ respectively.
Dashed lines (black, red, green): IAM results shown in Ref.~\cite{Wolff20} and
Ref.~\cite{PhysRevA.106.022813} for $\sigma_q$.
Dots (black, red, green): experimental values for $\sigma_q$ from fragmentation yields (Ref.~\cite{Wolff20}).
The results for $q=3$ are associated solely with $\rm H^+ + N^{2+}$ coincidences.
}
\label{fig:tcs}   
\end{center}
\end{figure}

In Fig.~\ref{fig:tcs} we show total cross sections for $\rm p-NH_3$ collisions over a wide range of energies.
On the one hand they allow to understand to what extent the discussion of shown DDCS is representative of the 
entire collision problem at a given impact energy. On the other hand the comparison allows us to discuss the relationship
with experimental total cross sections based on fragmentation yield measurements~\cite{Wolff20}, which were analyzed
recently within the IAM framework to compare them with supposedly similar target molecules, such as $\rm H_2O$ and $\rm CH_4$.

The data for $E_{\rm p} = 250 \ \rm keV$ in Fig.~\ref{fig:tcs} show how the net cross section is dominated by single ionization,
while  $\sigma_1+2\sigma_2+3\sigma_3$ is a lower bound to the net cross section, and the CTMC result is quite close to the data point of Ref.~\cite{Lynch1976}.
Together with the comparison of DDCS data shown in Fig.~\ref{fig:ddcs250} we may conclude that the CTMC-IEM presents a consistent picture
in the sense that about 40 \% of the net ionization cross sections comes from double (and higher) direct ionization processes.
Comparison with the experimental fragmentation data of Wolff~{\it et al.}, on the other hand, would imply that the net cross section is dominated
by $\sigma_1$ alone.

At higher energies the CTMC result for $\sum_{q=1}^3{q \, \sigma_q}$ falls barely below the experimental data.
This indicates that the energy dependence of the $\hbar=0$ semiclassical approach to the electron dynamics
based on the microcanonical distribution is incorrect at high energies on account of the sharp cut-off in the probability density
as a function of distance. For the atomic hydrogen H(1s) target this problem was identified and different attempts were made
to remedy the problem using other initial distributions than the microcanonical one~\cite{Eichenauer_1981,Cohen_1985,Schmidt_1990}.
Another reason is the lack of a dipole ionization mechanism in distant collisions~\cite{Reinhold_1993}.

When looking at the CTMC data for $\sigma_1(E_{\rm p})$ we notice that they are similar in shape to the experimental data of Ref.~\cite{Wolff20},
but lower by about one third. While at low collision energies $\sigma_2(E_{\rm p})$ can make up for this shortfall when comparing with the total (net)
cross section, it fails to do so at high energies due to the faster fall-off with $E_{\rm p}$.

The comparison with the IAM results given by the dashed black line (Refs.~\cite{Wolff20,PhysRevA.106.022813}) 
shows that the IAM is making a different prediction: its values for $\sigma_1$
match the experimental data at medium to high collision energies rather well. Only at lower $E_{\rm p}$ do the contributions from $\sigma_2$ start playing a more significant
role for the net cross section.

The double ionization data $\sigma_2$ highlight this discrepancy between the two theories and experiment. 
The quantum-mechanical IAM results show a steeper fall-off for the ratio $\sigma_2(E_{\rm p})/\sigma_1(E_{\rm p})$
than the present CTMC results (cf. Fig.~6 in Ref.~\cite{PhysRevA.106.022813}). For the IAM results the ratio is inversely proportional to $E_{\rm p} \log{(E_{\rm p})}$
at the higher energies, while the CTMC result follows some power law ($\sim E_{\rm p}^{-0.6}$).
The experimental data have been explained in Ref.~\cite{Wolff20} in terms of decay of satellite states, i.e., not at all consistent with a direct
multiple ionization picture.

The case of $q=3$ may illustrate limitations of the IEM approach for singly charged projectile ions. The CTMC cross section is rather large
and matches the experimental $\sigma_2$ values. This indicates that the IAM which uses atomic IEM cross sections and combines them geometrically 
has an advantage in this respect. The experimental data are from a single coincidence channel that corresponds to $q=3$, and
may be too low since two-proton coincidences together with a singly charged nitrogen atom cannot be recorded with the 
experimental set-up. Multiple proton coincidences were observed with a special coincidence technique in $\rm p-H_2O$ collisions
by Werner~{\it et al.}~\cite{Werner95}. Thus it is not unreasonable to assume that the experimental data of Ref.~\cite{Wolff20} are only a lower
bound for $\sigma_3$.

Presently, we cannot offer any resolution to this controversy except to re-iterate that the experimental fragmentation data
are inconsistent not only with the quantum-mechanical IAM approach, but also with the semiclassical CTMC method. Both theoretical approaches
take the molecular geometry fully into account.

\section{Conclusions}
\label{sec:conclusions}

We have presented a model calculation for fast proton-ammonia molecule collisions. An independent electron model is introduced
in analogy with previous works which involve the water molecule which from the point of view of MO energies has a 
comparable valence electron structure. This IEM was solved at the $\hbar=0$ level and doubly differential cross sections
were shown to be of overall competitive quality with approaches based on quantum scattering theory. These approaches follow
a scheme of linearly combined atomic orbitals, and effectively do orientation averaging before the collision process is considered. The present CTMC model
makes a reasonable prediction for the net ionization cross section.

The DDCS comparison with experiment and the mentioned quantum calculations is complemented by explicitly showing the
contributions from singly ionizing collision events, which demonstrates that multiple ionization does play an important role
within this IEM approach. Previously, a discrepancy with experiments for fragmentation yields as a function of energy was
reported, particularly, with respect to the differences with water and methane targets for which direct multiple ionization appears
to be the dominant contributor to double, and even triple ionization for $\rm CH_4$ and $\rm H_2O$ 
respectively~\cite{Wolff20,PhysRevA.106.022813}. The present work shows that the discrepancy persists in an IEM approach.
Since the experimental fragment collection may suffer from incomplete detection (e.g., simultaneous arrival of two protons),
one may argue that the problem deserves further experimental investigation.
If the sequence of similar targets $\rm H_2O, CH_4, NH_3$, on the other hand is confirmed to display such different behavior
in terms of multiple ionization, then the agreement with experimental net DDCS data obtained by various methods 
would seem rather fortuitous. A further investigation in this respect using a CDW-EIS approach, 
e.g. that of Ref.~\cite{Gulyas_2016}, which explicitly addressed multiple ionization in proton-water collisions, would
also provide more clarity and lead to some resolution.

\begin{acknowledgments}
We thank Hans J\"urgen L\"udde for many discussions.
Financial support from the Natural Sciences and Engineering Research Council of Canada (NSERC) 
(RGPIN-2017-05655 and RGPIN-2019-06305) is gratefully acknowledged. 
\end{acknowledgments}

\bibliography{ammonia}

\end{document}